\documentclass[a4paper,12pt]{article}

\usepackage{amsmath,amssymb,bm}
\usepackage[vcentermath,enableskew]{youngtab}

\usepackage{cite}

\usepackage{setspace}
\usepackage{fullpage}
\usepackage{graphicx}
\usepackage{amsfonts}

\usepackage[colorlinks=true]{hyperref}

\usepackage{color}

\definecolor{black}{rgb}{0,0,0}                                

\definecolor{blue}{rgb}{0,0,1}

\definecolor{green}{rgb}{0,1,0}

\definecolor{red}{rgb}{1,0,0}

\newcommand{\eq}{\begin{equation}}
\newcommand{\qe}{\end{equation}}
\newcommand{\non}{\nonumber}

\newcommand{\al}{\alpha}
\newcommand{\bt}{\beta}
\newcommand{\g}{\gamma}
\newcommand{\dl}{\delta}

\newcommand{\s}{\sigma}
\newcommand{\lm}{\lambda}
\newcommand{\avec}{\ensuremath{\bm{\alpha}}}
\newcommand{\bvec}{\ensuremath{\bm{\beta}}}
\newcommand{\dvec}{\ensuremath{\bm{\delta}}}
\newcommand{\ovec}{\ensuremath{\bm{\omega}}}
\newcommand{\rvec}{\ensuremath{\bm{\rho}}}
\newcommand{\svec}{\ensuremath{\bm{\sigma}}}
\newcommand{\tauvec}{\ensuremath{\bm{\tau}}}
\newcommand{\tvec}{\ensuremath{\bm{\theta}}}
\newcommand{\gvec}{\ensuremath{\bm{\gamma}}}

\newcommand{\mvec}{\ensuremath{\bm{\mu}}}
\newcommand{\nvec}{\ensuremath{\bm{\nu}}}
\newcommand{\evec}{\ensuremath{\bm{\eta}}}
\newcommand{\lvec}{\ensuremath{\bm{\lambda}}}
\newcommand{\vv}[1]{\ensuremath{\bm{#1}}}
\newcommand{\Avec}{\ensuremath{\bm{A}}}
\newcommand{\Bvec}{\ensuremath{\bm{B}}}
\newcommand{\Int}{\int_{\Gr_{2;n}}\!\!\!\!\! \Omega ~ }

\newcommand{\bM}{{\bf M}}
\newcommand{\bY}{{\bf Y}}

\newcommand{\bD}{{\bf D}}

\newcommand{\n}{\raise4pt \hbox{\tiny{${\bf n}$}}}
\newcommand{\nmo}{ \raise4pt \hbox{\tiny{$\bf~\! n-1 $ }} }
\newcommand{\one}{\raise4pt \hbox{ \hspace{-4mm} \tiny{ $\bf 1$} }}
\newcommand{\two}{\raise4pt \hbox{\tiny{$\bf 2$}}}
\newcommand{\Cdots}{\raise3pt \hbox{\small{$\cdots$}}}

\renewcommand{\P}{{\bf P}}
\newcommand{\calP}{{\mathcal{P}}}

\newcommand{\F}{\mathcal F}
\newcommand{\N}{\hat{\mathcal N}}
\newcommand{\A}{\hat{\mathcal A}}
\renewcommand{\L}{\hat{\mathcal L}}

\newcommand{\Lh}{\hat{L}}

\newcommand{\ttR}{{\tt R}}

\newcommand{\ad}[2]{(a^\dagger)_{#1}^{#2}}
\renewcommand{\a}[2]{a^{#1}_{#2}}

\newcommand{\Ad}{A^\dagger}

\newcommand{\CP}{\mathbb{C}\mathbb{P}}
\newcommand{\Gr}{\mathbb{G}\mathrm{r}}

\newcommand{\ket}[1]{\left| \! \right. #1\left. \! \right\rangle}
\newcommand{\bra}[1]{\left\langle \! \right. #1\left.\! \right| }

\def\tyng(#1){\hbox{\tiny$\yng(#1)$}}
\def\tyoung(#1){\hbox{\tiny$\young(#1)$}}

\title{Polarization Tensors for the fuzzy Grassmannians $\Gr^F_{2;n}$ and Grassmannian harmonics on $\Gr_{2;n}$}
\author{ Idrish Huet$^a$\footnote{idrish.huet@uni-jena.de}, Se\'an Murray$^b$\footnote{smury@stp.dias.ie}{ }\footnote{Present Address: Department of Computational and Systems Biology, John Innes Centre, Colney, Norwich, NR4 7UH, UK}  \\
\\
$^a$ {\it Theoretisch Physikalisches Institut,}\\
{\it Friedrich Schiller Universit\"at}\\
{\it Max-Wien-Platz 1}\\
{\it D-07743, Jena, Germany}
\\
\\
$^b$ {\it Centre for Cosmology, Particle Physics and Phenomenology (CP3)}\\
{\it Universit\'e Catholique de Louvain}\\
{\it Chemin du Cyclotron, 2},\\
{\it B-1348 Louvain-la-Neuve, Belgium}
}

\begin{document}

\maketitle

\begin{abstract}
We explicitly construct the eigenfunctions of the Laplacian for the fuzzy Grassmannian spaces $\Gr^F_{2;n}$. We calculate the spectrum and find it be a truncation of the 
continuum case. As a byproduct of our approach we find a novel expression for the Grassmannian harmonics in terms of Pl\"ucker coordinates which can be interpreted as free Schr\"odinger 
particle wave functions on $\Gr_{2;n}$.

\end{abstract}

\section{Introduction}

The complex Grassmannians, denoted here as $\Gr_{r;n}$ , are natural generalisations of the complex projective spaces. They are projective manifolds widely studied in 
algebraic geometry and have several applications throughout mathematical physics. Of particular interest is the family $\Gr_{2;n}$. This family of Grassmannians has 
remarkable significance to physics, e.g. it is known that $\Gr_{2;4}$ provides, through reality conditions, the compactified form of 4-dimensional $\mathbb{R}^{s,t}$ space-times of 
any signature. It was also shown by B. P.  Dolan and C. Nash in \cite{Dolan:2002af} that the zero modes of the Dirac operator on $\Gr_{2;5}$ reproduce the correct representations and 
charges of the fermionic sector of the standard model, up to multiplicites. Although the family $\Gr_{2;n}$ has been known for a long time, their fuzzy versions are relatively recent 
and a fuzzy Dirac operator on such Grassmannians would be desirable. To this end, an inital study of the fuzzy Laplacian and its eigenfunctions should be undertaken; this is the goal 
of the present work. Along the way we find a (as far as we know novel) construction of the harmonic functions on $\Gr_{2;n}$ in terms of Pl\"ucker coordinates. The Laplacian is a 
central geometrical object in classical and noncommutative differential geometry. In fact, it is part of the spectral triple definition of a noncommutative space \`a la Connes 
\cite{Connes:1994yd}. It has been know for some time that it encodes signifcant geometrical information \cite{Gilkey:2008}. 
In this context the eigenfunctions of the Laplacian are of fundamental importance for the harmonic analysis of a manifold. When properly normalised, they are interpreted in physics 
as the free Schr\"odinger particle's wave functions on the relevant manifold. This is the well known situation with the spherical harmonics and $S^2 = \CP^1$. We are interested in 
the fuzzy analogue of this construction for Grassmannians. When a fuzzy manifold $\mathcal{M}_F$ has a commutative version $\mathcal{M} = G/H$, with $G$ a simple and compact 
Lie group and $H$ a semisimple compact closed subgroup of $G$, the $G$-invariant Laplacian $\Delta_{\mathcal{M}}$ has a simple construction: by retaining the $G$-symmetry of the 
induced metric on $\mathcal{M}$ it can be shown that the associated $G$-invariant Laplacian is just the quadratic Casimir operator of $G$ formed with the right-invariant induced 
vector fields \cite{Ikeda:1978, Pilch:1984xx, elChami:2004}. Under these conditions the eigenfunctions of the Laplacian on $\mathcal{M}_F$ are given by the polarization tensors of a 
truncated family of irreducible representations of $G$. 

In this paper we focus our attention on the family of complex Grassmannians which have $G = U(n)$ and $H = U(2) \times U(n-2)$, i.e. $\Gr_{2;n}= U(n)/(U(2)\times U(n-2))$ 
and their fuzzy versions $\Gr^F_{2;n}$; we give their $U(n)$-invariant Laplacian and eigenfunctions in a Schwinger-Fock formalism, extending the one used for fuzzy complex projective 
spaces in \cite{Dolan:2006tx}. Already in \cite{Dolan:2001mi} the construction of the fuzzy Grassmannians $\Gr^F_{2;n}$ was given along with a star product and the corresponding 
coherent state map to functions, the relevant coherent states were constructed in \cite{Freidel:2010tt} and a Berezin-Toeplitz version was first given in \cite{BenHalima:2010}. The approach used 
here is different to that taken in \cite{Dolan:2001mi} in that it is based on composite pseudo-oscillators and a double Fock vacuum. This was first given for $\CP^n_F$ in 
\cite{Dolan:2006tx} and later generalised in \cite{Murray:2006pi} to flag and super-flag manifolds. In this context, we give the eigenfunctions of the Laplacians on $\Gr_{2;n}^F$ and 
$\Gr_{2;n}$ as an orthonormal basis under the standard trace and $L^2$-inner products respectively; we shall refer to the eigenfunctions of the fuzzy Laplacian simply as polarization 
tensors and the eigenfunctions of the commutative Laplacian as (Grassmannian) harmonics.

The paper is organised as follows: Section \ref{review} reviews the classical construction of the complex Grassmannians $\Gr_{r;n}$, introduces the Pl\"ucker coordinate system 
within the coset space realisation that we use for $\Gr_{2;n}$ and its fuzzy version $\Gr_{2;n}^F$. In section \ref{SectOsc} we introduce the composite (Schwinger) pseudo-oscillator 
construction of the fuzzy Pl\"ucker coordinates and the corresponding double Fock vacuum realisation of the function space of $\Gr^F_{2;n}$, the relevant $U(n)$ and $U(2)$ generators are 
given in terms of the fundamental oscillators. In this section we also introduce the Laplacian in terms of the left action derivatives of $su(n)$, we give also the representation content 
of the space of functions of $\Gr^F_{2;n}$ and define the polarization tensors that diagonalise the Laplacian. Section \ref{SectSpec} contains the calculation of the spectrum of the 
Laplacian. In section \ref{SectNorm} a recursive relation is found for the normalisation coefficients appearing in the polarization tensors in order to make them orthonormal under the 
trace inner product, we also give the relation between our construction and that of \cite{Dolan:2006tx} for the special case $\Gr^F_{2;3} = \CP^2$. Section \ref{SectHarm} presents the 
expression for the Grassmannian harmonics that is derived straightforwardly from our polarization tensors, and by use of results from previous sections, their normalisation in accordance 
to physics' conventions is fixed. Our discussion and conclusions are presented in section \ref{SectConc}. Finally, appendix \ref{SectP} contains an auxilary calculation used in section 
\ref{SectSpec} to compute the spectrum, while appendix \ref{SectQ} solves the recursion relation for the normalisation factors of the polarization tensors using the highest weight technique.
It also contains the proof of a combinatorial identity involving the symmetrizer used to solve the recursion relation.

\section{Review of $\Gr_{r;n}$} \label{review}

In this section, we review the standard construction of the complex Grassmannians $\Gr_{r;n}$ and introduce the Pl\"ucker coordinates, specializing to our case of interest 
$\Gr_{2;n}$. Further details may be found for instance in \cite{Chern:1999}.

Let $R \subset \mathbb{C}^n$ be a complex $r$-dimensional linear subspace, then we construct the maximal exterior power $\bigwedge^r R $, it is obviously a one 
dimensional linear subspace of $\bigwedge^r \mathbb{C}^n$, hence we may naturally identify $\bigwedge^r R $ with a point, $[R]$, in the associated projective space so that 
$[R] \in \mathbb{P}(\bigwedge^r \mathbb{C}^n) : = (\bigwedge^r \mathbb{C}^n \backslash \{ 0 \})/ \mathbb{C}^*$. A classical theorem of linear algebra implies the isomorphism 
$\bigwedge^r \mathbb{C}^n \cong \mathbb{C}^{\binom{n}{r}}$ and hence $\mathbb{P}(\bigwedge^r \mathbb{C}^n) \cong \CP^{\binom{n}{r} -1}$, we obtain the following embeddings:

\eq \nonumber
\Gr_{r;n} \hookrightarrow \CP^{\binom{n}{r} -1} \hookrightarrow \mathbb{C}^{\binom{n}{r}}
\qe 
The first arrow is the Pl\"ucker embedding, it is an injective mapping that sends the set of all $R$ subspaces to points in the complex projective plane. The coordinates given by the 
bijective map $R \mapsto [R]$ are the well known Pl\"ucker coordinates. To make the correspondence explicit we may associate to $R$ its volume $r$-form $\omega$ in a given basis 
$\{v^1, \ldots, v^r \}$ and expand it in a basis $\bigwedge^r \mathcal{B}$ of $\bigwedge^r \mathbb{C}^n$ with $ \mathcal{B}= \{ e^1, \ldots, e^n \}$ the basis for $\mathbb{C}^n$ :

\eq
\omega = v^1 \wedge v^2 \wedge \cdots \wedge v^r = \zeta_{\alpha_1 \cdots \alpha_r} e^{\alpha_1} \wedge \cdots \wedge e^{\alpha_r}, \quad \alpha_1, \ldots, \alpha_r = 1, \ldots, n
\qe
Here $\zeta_{\alpha_1 \cdots \alpha_r}$ are the Pl\"ucker coordinates associated to $R$, they give a representative of the equivalence class defined, as above, through 
multiplication with $\mathbb{C}^*$. The relation $\omega \wedge \omega = 0$ gives a set of quadratic constraints in the coordinates $\zeta$, usually called the Pl\"ucker equations 
\cite{Chern:1999}.  

For our purposes we will use the standard right coset space realisation of the complex Grassmannians

\eq 
\Gr_{r;n} = \frac{U(n)}{U(r) \times U(n-r)}~,
\qe
the embedding of the stability subgroup and the defining equivalence relation are

\eq
h \in U (r) \times U(n-r) \hookrightarrow  \left(
\begin{array}{cc}
U (r) & 0  \\
0     & U(n-r)
\end{array}
\right), \quad \quad u \sim u' = u h, \quad u \in U(n) ~.
\qe
From here it is easy to see that the complex dimension of $\Gr_{r;n}$ is $r(n-r)$. It is also clear that $\Gr_{r;n}= \Gr_{n-r;n}$ and $\Gr_{1;n} = \CP^{n-1}$.

We focus now on the case $r=2$. First we write a general element $u \in U(n)$ following the parametrization into orthonormal column vectors 
$\{ u^{\al}_{\bt} \in \mathbb{C}: ~ \al,\bt =1,\cdots n \}$ given in \cite{Dolan:2006tx} and introducing a rank 2 projector ${\bf p}$

\eq
u =\left(  \begin{array}{cccc}  u^1_1 & u^2_1 & \cdots & u^n_1 \\
                              u^1_2 & u^2_2 & \cdots & u^n_2  \\
                              \vdots &\vdots & \vdots & \vdots \\
                              u^1_n & u^2_n  & \cdots & u^{n}_n 
\end{array}
 \right), \quad \quad 
{\bf p } = \left(  \begin{array}{cccc}  1 & 0 & \cdots & 0 \\
                                        0 & 1 & \cdots & 0  \\
                                      \vdots & & \vdots & \\
                                        0 & 0 & \cdots & 0
\end{array}
 \right)~.
\qe
Unitarity implies the constraints $\bar{u}^{\al}_{\bt} u^{\bt}_{\g} = \bar{u}^{\bt}_{\g} u^{\al}_{\bt} = \dl^{\al}_{\g}$. To project onto $\Gr_{2;n}$ one acts on the right with 
the projector ${\bf p}$ so that $u \mapsto u {\bf p}$; only the two leftmost columns remain, leaving only components $u^i_{\al}$ with $i=1,2$. We obtain a set of Pl\"ucker coordinates for 
$\Gr_{2;n}$ through

\eq
\zeta_{\al \bt} = \frac{1}{\sqrt{2}} \epsilon_{ij} u^{i}_{\al}u^{j}_{\bt}~.
\qe
The normalisation has been chosen so that $\bar{\zeta}^{\al \bt}\zeta_{\al \bt} = 1$. By use of the antisymmetrization bracket $[~]$ in 
$\epsilon_{i[j}\epsilon_{kl]} u^{i}_{\al}u^{j}_{\bt}u^k_{\mu}u^{l}_\nu=0$ we obtain immediately the Pl\"ucker equations for $\Gr_{2;n}$

\eq
\zeta_{\al \bt} \zeta_{\mu \nu} + \zeta_{\al \mu} \zeta_{\nu \bt} + \zeta_{\al \nu} \zeta_{ \bt \mu }=0~.
\qe
\
In what follows we make use of these coordinates to give the fuzzy family of Grassmannians $\Gr_{2;n}^F$, first constructed in \cite{Dolan:2001mi}, and the explicit form of the 
polarization tensors of the corresponding $U(n)$ invariant Laplacian operating on functions.

\section{Fuzzy Grassmannians} \label{SectOsc}
To construct the noncommutative algebra describing $\Gr^F_{2;n}$, we proceed in the usual manner and replace the coordinates $\{ u^i_{\al}\}$ with bosonic oscillators \cite{Murray:2006pi}.
We introduce the Fock space $\F^T$ with vacuum $\ket{0}$ built from the bosonic creation and annihilation operators $\ad{\al}{i}$ and $\a{\al}{i}$ satisfying

\eq
\a{\al}{i}\ket{0}=0,\quad [\a{\al}{i},\,\ad{\bt}{j}]=\dl_\bt^\al \dl^j_i,~\qquad i,j=1, 2\quad \al,\bt=1,\ldots, n, \quad n\geq3~.
\qe
Then the composite creation and annihilation operators corresponding to the Pl\"ucker coordinates on $\Gr_{2;n}$ given in the last section are \cite{Murray:2006pi}

\eq 
\Ad_{\al \bt}= \frac{1}{\sqrt{2}}\epsilon_{ij}\ad{\al}{i}\ad{\bt}{j} \qquad \mathrm{and}\qquad A^{\al \bt}= \frac{1}{\sqrt{2}}\epsilon^{ij}\a{\al}{i}\a{\bt}{j} ~.
\qe
The only non-zero commutator is

\eq \label{commutAA} 
2[A^{\g\dl},\,\Ad_{\al\bt}]=\dl^\g_\al\hat{L}_\bt{}^\dl+\dl^\dl_\bt\hat{L}_\al{}^\g-\dl^\dl_\al\hat{L}_\bt{}^\g-\dl^\g_\bt\hat{L}_\al{}^\dl+
2\dl^\g_\al\dl^\dl_\bt-2\dl^\dl_\al\dl^\g_\bt,
\qe
where $\hat{L}_\al{}^\bt=\ad{\al}{i}\a{\bt}{i}$ are the generators of $U(n)$ as can be seen from

\eq
[\Lh^{~\al}_{\bt}, \Lh^{~\g}_{\dl} ] = \Lh^{~\g}_{\bt}\delta^{\al}_{\dl} - \Lh^{~\al}_{\dl} \delta^{\g}_{\bt}~,
\qe
they also satisfy the following commutator with the creation operators

\eq 
[\hat{L}_\al{}^\bt,\,\Ad_{\mu\nu}]=\dl^\bt_\mu \Ad_{\al\nu}+\dl^\bt_\nu \Ad_{\mu\al}\label{Jcommutator}~.
\qe
Hence these composite operators generate a $U(n)$-invariant subspace, $\F \subset \F^T$. Fixing the number of composite creation operators, we denote with $\F_L$ the space of 
states with a fixed number, $L$, of composite creation operators:

\eq
\F_L = \mbox{Span}_{\mathbb{C}}\{ \Ad_{\al_1\bt_1}\cdots \Ad_{\al_L\bt_L}\ket{0} \} ~.\label{states}
\qe
The number operators

\eq \N_i=\ad{\al}{i}\a{\al}{i}\qquad\mathrm{for~any~fixed~}i 
\qe
clearly take the value $L$ on this subspace, and so on $\F_L$ we write $\N=\N_i$.

It is straightforward to check that the operators $\Ad_{\al\bt}$ satisfy the following relation

\eq
[\hat{J}^i{}_j,\,\Ad_{\al\bt}]=\dl^i{}_j\Ad_{\al\bt} 
\qe
with the $U(2)$ generators $\hat{J}^i{}_j=\ad{\al}{i}\a{\al}{j}$. Therefore, states of $\F_L$ are $U(2)$ singlets 

\eq
\hat{J}^i{}_j \Ad_{\al_1 \bt_1}\cdots\Ad_{\al_L \bt_L}\ket{0}=\dl^i{}_j \N\Ad_{\al_1 \bt_1}\cdots\Ad_{\al_L \bt_L}\ket{0}~.\label{internal U(2) invariance}
\qe
We also find that $\Ad_{\al\bt}A^{\al\bt}=\N(\N+1)$ on $\F_L$. It is convenient to normalise the states in $\F_L$. To this end, observe that $\F_L$ carries the $\overbrace{\tyng(5,5)}^L$ 
representations of $su(n)$ and that therefore

\eq 
\label{bP} \bra{0}A^{\mu_1\nu_1}\cdots A^{\mu_L\nu_L}\Ad_{\al_1 \bt_1}\cdots\Ad_{\al_L \bt_L}\ket{0}=\alpha(L) \P_{\avec,\,\bvec}^{\mvec,\,\nvec}~,
\qe
where $\P_{\avec,\,\bvec}^{\mvec,\,\nvec}=\P^{\mu_1\cdots\mu_L,\,\nu_1\cdots\nu_L}_{\al_1\cdots\al_L,\,\bt_1\cdots\bt_L}$ is the projector of rank $2L$ tensors 
$T_{\mu_1\cdots\mu_L\nu_1\cdots\nu_L}$ to tensors with the index structure in (\ref{states}) and $\alpha (L)$ is a constant of proportionality. Since the trace of $\P$ is just the 
dimension of the $su(n)$ irreducible representation given above, namely

\eq
d_n(L)=\mathrm{dim}_{su(n)} \overbrace{\tyng(5,5)}^L = \frac{(L+n-1)!(L+n-2)!}{L!(L+1)!(n-1)!(n-2)!}
\qe
and using

\eq \label{AAd} 
A^{\al\bt}\Ad_{\al\bt}=(\N+n)(\N+n-1)~,
\qe
we find $\alpha(L)=L!(L+1)!$. We shall then define states $\ket{\avec,\,\bvec}$ with the normalisation\footnote{Our convention will be to always write indices in the order that they 
appear from left to right so that \eq \bra{\avec,\,\bvec}=\bra{0}A^{\al_1\bt_1}\cdots A^{\al_L\bt_L}\frac{1}{\sqrt{L!(L+1)!}} \qe}

\eq 
\ket{\avec,\,\bvec}=\frac{1}{\sqrt{L!(L+1)!}}\Ad_{\al_1\bt_1}\cdots \Ad_{\al_L\bt_L}\ket{0}\label{normedstates}~. 
\qe
The generators of $su(n)$ are given by

\eq 
\hat{L}_a=\ad{\al}{i} \frac{(\lambda_a)^\al_\bt}{2} \a{\bt}{i}~,
\qe
where $\lambda_a$ are the Murray-Gellmann matrices for $su(n)$. It is not difficult to see that

\eq 
\hat{L}^2 |_{\F_L}  =\hat{L}_a\hat{L}_a|_{\F_L}=\frac{n-2}{n}\N(\N+n)|_{\F_L}.
\qe
Let us now consider the finite dimensional algebra $\A_L$ given by a pairing of vectors from $\F_L$ and its dual $\F^*_L$ and having basis elements 
$\ket{\avec,\,\bvec}\bra{\mvec,\,\nvec}$

\eq 
\A_L=\F_L\otimes\F^*_L=\mathrm{span}\{\ket{\avec,\,\bvec}\bra{\mvec,\,\nvec}\}~.
\qe
It is isomorphic to the algebra of $d_n(L)$-dimensional matrices. Observe that the resolution of the identity on $\F_L$ is given by

\eq \label{resolution}
\ket{\avec,\,\bvec}\bra{\avec,\,\bvec}=\bm{1}~,
\qe
where a sum over repeated indices is implied.

The left action of $su(n)$ on an element $\bM \in \A_L$ is given by

\eq 
\L_a\bM=\hat{L}_a\bM-\bM\hat{L}_a~.
\qe
It is not difficult to see that $\L_a \bm{1}=0$, which is what we would expect for the analogue of a constant function.

\section{Spectrum of the fuzzy Laplacian} \label{SectSpec}

The Laplacian is given by the quadratic Casimir operator\footnote{The $su(n)$ and $u(n)$ quadratic Casimir operators built from left actions coincide in $\A_L$.}, so we have

\begin{align}
\L^2\bM&=\L_a\L_a\bM=\hat{L}^2\bM+\bM\hat{L}^2-2\hat{L}_a\bM\hat{L}_a\\
&=\frac{n-2}{n}\N(\N+n)\bM+\frac{n-2}{n}\bM\N(\N+n)+\frac{4}{n}\N\bM\N-\ad{\al}{i}\a{\bt}{i}\bM\ad{\bt}{j}\a{\al}{j}\\
&=2L(L+n-2)\bM-\hat{L}_{\al}{}^{\bt}\bM\hat{L}_{\bt}{}^\al~.\label{laplacian} 
\end{align}
Similarily to the case of $S^2_F=\CP^1_F$, the algebra of functions, $\A_L$, is spanned by polarization tensors according to a decomposition into irreducible representations of $su(n)$. 

\noindent Since $\F_L$ and $\F^*_L$ carry the representations $\overbrace{\tyng(4,4)}^L$ and $n-2\left\{\vphantom{\tyng(4,4,4,4)}\right.\overbrace{\tyng(4,4,4,4)}^L$ respectively, 
the relevant group theory decomposition is 

\eq
n-2\left\{\vphantom{\tyng(4,4,4,4)}\right.\overbrace{\tyng(4,4,4,4)}^L~\otimes~\overbrace{\tyng(4,4)}^L~.
\qe
For simplicity, we denote the anti-fundamental representation by one barred box

\eq
\overline{\tyoung(\ )}=n-1\left\{\tyoung(\ ,\ ,\ ,\ ) \right.~,
\qe
so that, for example, tensors transforming in the adjoint representation

\eq 
\overline{\tyng(1)}\hspace{-0.1mm}\tyng(1)=n-1\bigg\{ \tyoung(\ \ ,\ ,\ ,\ ) 
\qe
naturally have two indices, which have their contraction removed. We can now write down the decomposition:

\begin{equation}
\overbrace{\overline{\tyng(4,4)}}^{L}~\otimes~\overbrace{\tyng(4,4)}^L~\ =\ ~
\mathbf{1}~\oplus~\overline{\tyng(1)}\hspace{-0.1mm}\tyng(1)~\oplus~\overline{\tyng(1,1)}\hspace{-0.1mm}\tyng(1,1)~ \oplus~\overline{\tyng(2)}\hspace{-0.1mm}\tyng(2)~
\oplus~\overline{\tyoung(\ \ ,:\ )}\hspace{-0.1mm}\tyng(2,1)~\oplus~\overline{\tyng(2,2)}\hspace{-0.1mm}\tyng(2,2)~\oplus \ldots
\end{equation}
This expression is more easily written in terms of Dynkin indices as

\eq 
(0,\cdots,0,L,0)\otimes(0,L,0,\cdots,0)=\bigoplus_{l=0}^L\bigoplus_{p=0}^{L-l}~ (p,l,0,\cdots,0,l,p)~.
\qe
We shall write the polarization tensors corresponding to these representations as

\eq\label{Dharm}
\bD_{\avec_l,\bvec_l;\gvec_p}^{\mvec_l,\nvec_l;\evec_p}=\frac{1}{\sqrt{Q_n(l,p,L)}} \calP_{(\avec_l,\bvec_l;\gvec_p)(\avec_l',\bvec_l';\gvec_p')}^{(\mvec_l,\nvec_l;\evec_p)
(\mvec_l',\nvec_l';\evec_p')}\ket{\mvec_l'\evec'_p \vv{\theta} ,\nvec'_l\dvec_p \vv{\rho}}\bra{\avec'_l\gvec'_p \vv{\theta},\bvec'_l\dvec_p\vv{\rho}}~,
\qe
where $Q_n(l,p,L)$ is a normalisation factor and $\calP_{(\avec_l,\bvec_l;\gvec_p)(\avec'_l,\bvec'_l;\gvec'_p)}^{(\mvec_l,\nvec_l;\evec_p)(\mvec'_l,\nvec'_l;\evec'_p)}$ is the projector 
onto the representation $(p,l,0,\cdots,0,l,p)$ - it symmetrizes both the $\gvec$ and $\evec$ indices and removes all contractions between raised and lowered indices. We have also dropped 
an $L-p-l$ subscript on $\vv{\theta}$ and $\vv{\rho}$ for clarity.

We show in appendix \ref{SectP} that when $l+p=L$, the final term of the Laplacian (\ref{laplacian}) is given by

\eq \label{finallaplacian}
\hat{L}_{\al}{}^\bt\bD_{\avec_l,\bvec_l;\gvec_p}^{\mvec_l,\nvec_l;\evec_p}\hat{L}_{\bt}{}^\al=p(2L+n-p-3)\bD_{\avec_l,\bvec_l;\gvec_p}^{\mvec_l,\nvec_l;\evec_p}
\qe
and so from (\ref{laplacian})

\begin{align} 
\L^2\bD_{\avec_l,\bvec_l;\gvec_p}^{\mvec_l,\nvec_l;\evec_p}&=(2(l+p)^2+2(n-p-2)l+(n-p-1)p)\bD_{\avec_l,\bvec_l;\gvec_p}^{\mvec_l,\nvec_l;\evec_p}\non\\
&=C_n(l,p)\bD_{\avec_l,\bvec_l;\gvec_p}^{\mvec_l,\nvec_l;\evec_p} \label{Lapeigen}
\end{align}
where $C_n(l,p)$ is the eigenvalue of the second order Casimir operator on the $su(n)$ representation $(p,l,0,\cdots,0,l,p)$\footnote{The Grassmannian obtained when $n=3$ is 
isomorphic to $\mathbb{CP}^2$ \cite{Murray:2006pi}. In this case, we know that the appropriate representations are $(p,0,0,\cdots,0,0,p)$, evidenced by the correct spectrum 
$C_3(0,p)=p(p+2)$.}. After a short calculation, it can be seen that
\eq \L^2(\Ad_{\al\bt}\bM A^{\al\bt})=\Ad_{\al\bt}(\L^2\bM)A^{\al\bt}
\qe
for any $\bM\in \A_L$ and so the eigenvalues of the Laplacian are independent of $L$; they depend only on $l$ and $p$. Therefore the above eigenvalue equation holds for any $l$ and 
$p$ such that $0\leq l \leq L$, $0\leq l+p \leq L$, the spectrum of the Laplacian on $\Gr^F_{2;n}$ is then $\{C_n (l,p)\}$ with $p,l$ obeying the forementioned constraints. 
Therefore the fuzzy spectrum is identical to that of the continuum Laplacian on $\Gr_{2;n}$ up to a cut-off $L$ \cite{Pilch:1984xx}.

\section{Normalisation of Polarization Tensors} \label{SectNorm}

In this section we compute the normalisation factor $Q_n(l,p,L)$ of the polarization tensors given by (\ref{Dharm}) so that they satisfy the orthonormality condition

\eq \label{OrthonormalD}
Tr ( \bD_{\avec_l,\bvec_l;\gvec_p}^{\mvec_l,\nvec_l;\evec_p}\bD_{\avec_l',\bvec_l';\gvec_p'}^{\mvec_l',\nvec_l';\evec_p'})= 
\calP_{(\avec_l,\bvec_l;\gvec_p)(\avec_l',\bvec_l';\gvec_p')}^{(\mvec_l,\nvec_l;\evec_p)(\mvec_l',\nvec_l';\evec_p')}~.
\qe
First we give some useful identities used in the computation:

\begin{align} A^{\g \delta} \Ad_{\al \bt} &=\frac{1}{2}\left( 2\delta^{\g \dl}_{\al \bt} + \delta^{[\g}_{\al}\Lh^{~\dl ]}_{\bt} + 
2 \delta^{[\dl}_{\bt}\Lh^{~\g ]}_{\al} + \Lh^{~[\g}_{\al}\Lh^{~\dl ]}_{\bt}\right)\non\\
&=\frac{1}{2}\left(\hat{L}_\al{}^\g +\dl_\al^\g\right)\left(\hat{L}_\bt{}^\dl+2\dl_\bt^\dl\right)-\g \leftrightarrow \dl\non\\
&= \frac{1}{2}\left(\hat{L}_\al{}^\g +2\dl_\al^\g\right)\left(\hat{L}_\bt{}^\dl+\dl_\bt^\dl\right)-\alpha \leftrightarrow \beta~~,\label{AAdaggerexpansion}\\
\hat{L}_\al{}^\bt\hat{L}_\dl{}^\al&=\hat{J}^i{}_j \ad{\dl}{j} a^\bt_i+2\N\dl^\bt_\dl-2\hat{L}_\dl{}^\bt=\ad{\dl}{j} a^\bt_i\hat{J}^i{}_j +
2\N\dl^\bt_\dl-2\hat{L}_\dl{}^\bt ~~ ,\label{L/Lexpansion}\\
\hat{L}_\dl{}^\al\hat{L}_\al{}^\bt&=\hat{J}^i{}_j \ad{\dl}{j} a^\bt_i+(n-2)\hat{L}_\dl{}^\bt= \ad{\dl}{j} a^\bt_i\hat{J}^i{}_j+(n-2)\hat{L}_\dl{}^\bt~~.\label{LLexpansion}
\end{align}
Where $\delta^{\g \dl}_{\al \bt} = \delta^{\g}_{\al} \delta^{\dl}_{\bt} - \delta^{\g}_{\bt} \delta^{\dl}_{\al}$ and $[~]$ indicates anti-symmetrization without additional prefactors. 
We aim to calculate the normalisation factor $Q_n(l,p,L)$, to this end notice that we can use cyclicity of the trace to gain some information:

\begin{align} 
Tr\left(\bD_L A^{\al\bt}\Ad_{\g\dl}\bD'_L A^{\g\dl} \Ad_{\al\bt}\right)&=Tr\left(\Ad_{\al\bt}\bD_L A^{\al\bt}\Ad_{\g\dl}\bD'_L A^{\g\dl} \right)\non\\ \label{Rec1}
&=(L+1)^2(L+2)^2\frac{Q_n(l,p,L+1)}{Q_n(l,p,L)} Tr\left(\bD_{L+1}\bD'_{L+1}\right)
\end{align}
where $\bD_L$ is an obvious shorthand. Representation theory tells us that $Tr\left(\bD_{L}\bD'_{L}\right)$ is proportional to the projector 
$\calP_{(\avec_l,\bvec_l;\gvec_p)(\avec_l',\bvec_l';\gvec_p')}^{(\mvec_l,\nvec_l;\evec_p)(\mvec_l',\nvec_l';\evec_p')}$ if the irreducible representations corresponding to 
$\bD_L$ and $\bD'_L$ are the same and vanishes otherwise. Therefore, we can calculate $A^{\al\bt}\Ad_{\g\dl}\bD_L A^{\g\dl} \Ad_{\al\bt}$ to obtain a recurrence relation in $L$.

Using equations (\ref{AAdaggerexpansion}), (\ref{L/Lexpansion}), (\ref{LLexpansion}) and (\ref{internal U(2) invariance}) and after a long calculation we find that

\begin{align} \label{AAMAA}
A^{\al\bt}\Ad_{\g\dl}\bM A^{\g\dl} \Ad_{\al\bt}&=\frac{1}{2}\left( \hat{L}_\al{}^\g\hat{L}_\bt{}^\dl \bM\hat{L}_\dl{}^\bt\hat{L}_\g{}^\al -
\hat{L}_\al{}^\g\hat{L}_\bt{}^\dl \bM\hat{L}_\dl{}^\al\hat{L}_\g{}^\bt \right.\non\\&+\left. (6L+n+3)\hat{L}_\bt{}^\al \bM\hat{L}_\al{}^\bt + 4(2L+n)(2L+n-1) \right) \bM
\end{align}
for any $\bM \in \A_L$. 

Acting now on a given polarization tensor $\bD$, using the expression for general $L$, from (\ref{laplacian}) and (\ref{finallaplacian}): 
$\Lh^{~\bt}_{\al} \bD \Lh^{~\al}_{\bt} = [2L(L +n-2)-C_n (l,p)]\bD$, and so only $\hat{L}_\al{}^\g\hat{L}_\bt{}^\dl \bD\hat{L}_\dl{}^\al\hat{L}_\g{}^\bt$ is left. Using 
(\ref{Jcommutator}) and equations (\ref{contract1}) and (\ref{contract2}) from appendix \ref{SectP} repeatedly we obtain 

\begin{eqnarray} \non 
\Lh^{~\g}_{\al} \Lh^{~\dl}_{\bt} \bD \Lh^{~\al}_{\dl} \Lh^{~\bt}_{\g} &=& \Big[ \big( (p+q)^2 + q^2 +p \big)(\N + \N^\ttR )^2   \\ \non
&+& \big( q(n-2)(4q-4+p) - 4q^3 +p(n-2q-p-3)(2p + 3q + 1) \\ \non
& + & p(3 + pq -n-q)\big)(\N + \N^\ttR ) \\ 
 & + &  2q(n-q-2)[(q-2)(n-2)-q^2] + p(n-2q-p-3) \\ \non
 & \times & [(p+2q)(n-q-3-p) + pq + 3 - n] \Big] \bD ~.
\end{eqnarray}
Where we defined $q= L-p-l$ for convenience. The superscript ${}^\ttR$ indicates that the operator acts upon the algebra on the right. Substituting this expression back into 
(\ref{AAMAA}) with $\bM = \bD$ yields finally

\eq \label{AADAA}
A^{\al\bt}\Ad_{\g\dl}\bD A^{\g\dl} \Ad_{\al\bt}=(L-l+2)(L-l-p+1)(L+l+n-1)(L+n+l+p)\bD~. 
\qe
One may check (\ref{AADAA}) against the known result for $\mathbb{CP}^2$. From \cite{Dolan:2006tx} and in the notation of that paper, one can easily obtain the following 
result

\eq \label{AAYAA} 
\tilde{A}^{\al}\tilde{A}^\dagger_{\bt}\bY_{\mvec}{}^{\bm{\nu}} \tilde{A}^{\bt} \tilde{A}^\dagger_{\al}=(L+2)^2(L-l+1)(L+l+3)\bY_{\mvec}{}^{\bm{\nu}}~, 
\qe
which matches with the result with $n=3$, $l=0$ and $p \rightarrow l$. The relation between our two-index oscillators and the single index-index ones in the case of 
$\mathbb{CP}^2 = \Gr_{2;3}$ is given by the correspondence

\eq
\frac{1}{\sqrt{2}} \epsilon_{\al \bt \g} A^{\bt \g} \longrightarrow \tilde{A}^{\al}
\qe
as can be easily seen from (\ref{AADAA}) and (\ref{AAYAA}) or computing the commutation relations (\ref{commutAA}) in terms of $\tilde{A}^{\al}$ and reffering back to \cite{Dolan:2006tx}.

Using now equation (\ref{Rec1}) we find the recurrence relation in $L$ 

\eq
Q_n (l,p,L) = \frac{(L-l+1)(L-l-p)(L+l+n-2)(L+n+l+p-1)}{L^2 (L+1)^2} Q_n (l,p,L-1) 
\qe
valid down to the minimum value $L = l+p$. A short calculation yields

\begin{eqnarray} 
Q_n (l,p,L) &=& \frac{(L-l+1)! (L-l-p)!(L+l+n-2)!(L+n+l+p-1)!}{(p+1)!(2l+p+n-2)!(2l+2p+n-1)!} \\ \non & \times & \left(\frac{(l+p)!(l+p+1)!}{L!(L+1)!}\right)^2 Q_n (l,p,l+p)
\end{eqnarray}
the remaining computation of the normalisation factor is completed with the result for $Q_n (l,p,l+p)$ found in appendix \ref{SectQ}.

\section{Grassmannian harmonics on $\Gr_{2;n}$} \label{SectHarm}

The eigenfunctions of the Laplacian can now be given in terms of Pl\"ucker coordinates, they are the commutative limit of (\ref{Dharm}), namely:

\eq \label{Yharm}
\bY^{\mvec_l, \nvec_l ; \evec_p}_{\avec_l, \bvec_l ; \gvec_p} = \frac{1}{\sqrt{q_n (l,p)}} 
\calP^{(\mvec_l,\nvec_l ; \evec_p)(\mvec_l',\nvec_l' ; \evec_p')}_{(\avec_l,\bvec_l ;\gvec_p)(\avec_l',\bvec_l' ;\gvec_p')} \zeta_{\mvec_l' \nvec_l'}\zeta_{\evec_p' \dvec_p}
\bar{\zeta}^{\avec_l' \bvec_l'}\bar{\zeta}^{\gvec_p' \dvec_p}
\qe
here we have used the shorthand $\zeta_{\mvec_k \nvec_k} = \zeta_{\mu_1 \nu_1} \cdots \zeta_{\mu_k \nu_k} $ for $k = l,p$. The normalisation factor will be chosen so that the 
orthonormality relation 

\eq \label{Yorth}
\Int  \bY^{\mvec_l, \nvec_l ; \evec_p}_{\avec_l, \bvec_l ; \gvec_p} \bY^{\mvec_l', \nvec_l' ; \evec_p'}_{\avec_l', \bvec_l' ; \gvec_p'} = 
\calP^{(\mvec_l,\nvec_l ; \evec_p)(\mvec_l',\nvec_l' ; \evec_p')}_{(\avec_l,\bvec_l ;\gvec_p)(\avec_l',\bvec_l' ;\gvec_p')}
\qe
is satisfied for the $U(n)$-invariant measure induced on $\Gr_{2;n}$, $\Omega$. We remark that $\bY^{\mvec_l, \nvec_l ; \evec_p}_{\avec_l, \bvec_l ; \gvec_p}$ are essentially the wave 
functions of the Schr\"odinger free particle moving on $\Gr_{2;n}$, they generalise the spherical harmonics to the mentioned family of Grassmannians. In what follows we compute 
$q_n(l,p)$, the normalisation factor required in these wave functions, it will be shown that

\begin{eqnarray} \label{qn}
q_3 (l,p)&=& \frac{\mbox{Vol}(\Gr_{2;n}) ((l+p)! (l+p+1)!)^2 }{2^{2l+2p-1} (2l+2p+1)! (2l+2p+2)!} \\ \nonumber
q_n (l,p) &=& \frac{\mbox{Vol}(\Gr_{2;n}) (l! p! (l+p+1)!)^2 (n-1)!(n-2)! }{4^{l+p} (2l+2p+n-2)! (2l+2p+n-1)!} \binom{2l+p+n-3}{p},  \quad \quad n \geq 4 ~.
\end{eqnarray}
The only integral to be evaluated in (\ref{Yorth}) is

\eq \label{int1} 
J= \Int \zeta_{\rvec_l \svec_l } \zeta_{\rvec_l' \svec_l'} \zeta_{\tvec_p \dvec_p }\zeta_{\tvec_p' \dvec_p'} \bar{\zeta}^{\lvec_l \tauvec_l}\bar{\zeta}^{\lvec_l' \tauvec_l'}
\bar{\zeta}^{\ovec_p \dvec_p} \bar{\zeta}^{\ovec_p' \dvec_p'}~,
\qe
and this can be evaluated using representation theory. If we conceive such integral as an endomorphism of the representation space $\F_{2(l+p)}$ then the 
$U(n)$-invariance of the measure and Schur's lemma imply that the integral must be proportional to (\ref{bP}) with $L=2(l+p)$ and the appropriate indices. By taking traces over 
$\F_{2(l+p)}$ we obtain the proportionality constant and find 

\eq \label{PPP}
J = \frac{\mbox{Vol}(\Gr_{2;n})}{d_n( 2(l+p) )}
\P^{\lvec_l \lvec_l' \ovec_p \ovec_p',\, \tauvec_l \tauvec_l' \dvec_p \dvec_p'}_{\rvec_l \rvec_l' \tvec_p \tvec_p' , \, \svec_l \svec_l' \dvec_p \dvec_p'} ~.
\qe
Now we use (\ref{bP}) and perform the following transformations

\begin{eqnarray} \label{alP}
\al (2(l+p))\P^{\lvec_l \lvec_l' \ovec_p \ovec_p',\, \tauvec_l \tauvec_l' \dvec_p \dvec_p'}_{\rvec_l \rvec_l' \tvec_p \tvec_p' , \, \svec_l \svec_l' \dvec_p \dvec_p'}&=& 
\bra{0} A^{\lvec_l \tauvec_l}A^{\lvec_l' \tauvec_l'}A^{\ovec_p \dvec_p}A^{\ovec_p' \dvec_p'} A^{\dagger}_{\rvec_l \svec_l}A^{\dagger}_{\rvec_l' \svec_l'} 
A^{\dagger}_{\tvec_p \dvec_p} A^{\dagger}_{\tvec_p' \dvec_p'} \ket{0} \\ \nonumber
&=& \bra{0} A^{\lvec_l \tauvec_l}A^{\ovec_p \dvec_p}A^{\dagger}_{\rvec_l' \svec_l'}A^{\dagger}_{\tvec_p' \dvec_p'}A^{\lvec_l' \tauvec_l'}A^{\ovec_p' \dvec_p'}
 A^{\dagger}_{\rvec_l \svec_l} A^{\dagger}_{\tvec_p \dvec_p}  \ket{0} + \sigma \\ \nonumber
&=& \bra{0} A^{\lvec_l \tauvec_l}A^{\ovec_p \dvec_p}A^{\dagger}_{\rvec_l' \svec_l'}A^{\dagger}_{\tvec_p' \dvec_p'} {\bf 1}_{\F} A^{\lvec_l' \tauvec_l'}A^{\ovec_p' \dvec_p'}
 A^{\dagger}_{\rvec_l \svec_l} A^{\dagger}_{\tvec_p \dvec_p}  \ket{0} + \sigma~.
\end{eqnarray}
It is obvious that the term $\sigma$ contains only terms that will vanish upon contractions in (\ref{Yorth}). It should be clear that by $A^{\lvec_l \tauvec_l}$ we mean 
$A^{\lm_1 \tau_1} \cdots A^{\lm_l \tau_l} $ and so on.  We inserted the identity acting on the Fock space, after (\ref{resolution}) we have 
${\bf 1}_{\F} = \bigoplus_{L=0}^{\infty} | \avec ,\bvec \rangle \langle \avec, \bvec |$.

 Observe that only the term $L=0$ contributes, therefore (\ref{alP}) becomes

\begin{eqnarray} \nonumber
& & \bra{0} A^{\lvec_l \tauvec_l}A^{\ovec_p \dvec_p}A^{\dagger}_{\rvec_l' \svec_l'}A^{\dagger}_{\tvec_p' \dvec_p'} \ket{0} \bra{0}  A^{\lvec_l' \tauvec_l'}A^{\ovec_p' \dvec_p'}
 A^{\dagger}_{\rvec_l \svec_l} A^{\dagger}_{\tvec_p \dvec_p}  \ket{0} + \sigma \\ 
& & \quad= ~ \al (l+p)^2 \P^{\lvec_l \ovec_p, \, \tauvec_l \dvec_p}_{\rvec_l' \tvec_p' , \, \svec_l' \dvec_p'} 
\P^{\lvec_l' \ovec_p', \, \tauvec_l' \dvec_p'}_{\rvec_l \tvec_p , \, \svec_l \dvec_p} + \sigma
\end{eqnarray}
With these observations can then rewrite the l.h.s of (\ref{Yorth}) as

\eq \label{PPP2}
\eta \calP^{(\mvec_l,\nvec_l ; \evec_p)(\rvec_l,\svec_l ; \tvec_p)}_{(\avec_l,\bvec_l ;\gvec_p)(\lvec_l,\tauvec_l ;\ovec_p)}
\calP^{(\mvec_l',\nvec_l' ; \evec_p')(\rvec_l',\svec_l' ; \tvec_p')}_{(\avec_l',\bvec_l' ;\gvec_p')(\lvec_l',\tauvec_l' ;\ovec_p')}
\P^{\lvec_l \ovec_p, \, \tauvec_l \dvec_p}_{\rvec_l' \tvec_p' , \, \svec_l' \dvec_p'} \P^{\lvec_l' \ovec_p', \, \tauvec_l' \dvec_p'}_{\rvec_l \tvec_p , \, \svec_l \dvec_p}
\qe
where we set $\eta = \frac{\al(l+p)^2 \mbox{Vol}(\Gr_{n;2})}{\al (2(l+p))q_n(l,p)d_n(2(l+p))}$. We can then evaluate (\ref{PPP2}) from (\ref{OrthonormalD}) and (\ref{bP}) by writing 
out the polarization tensors $\bD_L$ with $L=l+p$, the final result is that (\ref{PPP2}) reduces to 

\eq
\eta Q_n(l,p,l+p) \calP^{(\mvec_l,\nvec_l ; \evec_p)(\mvec_l',\nvec_l' ; \evec_p')}_{(\avec_l,\bvec_l ;\gvec_p)(\avec_l',\bvec_l' ;\gvec_p')}~,
\qe
condition (\ref{Yorth}) implies $\eta Q_n (l,p,l+p)=1$, which together with the result for $Q_3 (l,p,l+p)$ and (\ref{Qminimal}) found in appendix \ref{SectQ} gives (\ref{qn}) as 
promised.

Finally we would like to quote from \cite{Fujii:2001qga} the general result for the volume of unitary Grassmannians that corresponds to our normalisation of the Pl\"ucker coordinates, and 
for our case at hand:

\eq
\mbox{Vol}(\Gr_{r;n}) = \pi^{r (n-r)}\prod_{j=1}^r \frac{(j-1)!}{(n-j)!}, \quad \mbox{Vol}(\Gr_{2;n}) = \frac{\pi^{2(n-2)}}{(n-1)!(n-2)!}~.
\qe

\section{Concluding remarks} \label{SectConc}

In this paper we have used a Schwinger-Fock type construction of the family of fuzzy Grassmannians $\Gr_{n;2}^F$ to compute the eigenfunctions and spectrum of their Laplacian and that of 
$\Gr_{2;n}$. In particular we used two-index pseudo-oscillators which correspond to the fuzzy Pl\"ucker coordinates as the generators of the fuzzy function algebra acting on a double 
Fock vacuum, the fuzzy function algebra is found to be a $U(2)$ singlet, with the relevant $U(2)$ appearing in the isotropy subgroup of $U(n)$.

The $su(n)$ left action derivatives on $\Gr_{2;n}^F$, $\L_a$, were also given and the corresponding $U(n)$-invariant Laplacian built from them. Using representation theory the 
polarization tensors (eigenfunctions of the fuzzy Laplacian) were defined on the fuzzy Grassmannians $\Gr_{n;2}^F$ and the corresponding spectrum was calculated. The special case 
$n=3$, for which $\Gr_{3;2} = \CP^2$, was shown to be in agreement with previous results found already in \cite{Dolan:2006tx}; the spectrum and the algebra of fuzzy functions are 
recovered.

By taking the commutative limit we were able to give an explicit expression for the Grassmannian harmonics on $\Gr_{2;n}$ in the Pl\"ucker coordinate system. To our knowledge this 
construction of the Grassmannian harmonics is new. Using the highest weight technique we normalised both the Grassmannian harmonics and polarization tensors, following conventions 
from physics, under the trace and $L^2$-norm.

Further work along these lines suggests the study of right action covariant derivatives and the construction of a fuzzy Dirac operator and the corresponding eigenspinors on both 
$\Gr_{2;n}^F$ and $\Gr_{2;n}$ should be possible, at least in the case $n$ even when they admit a spin structure. The study of universal covariant derivatives and the connection between 
this Schwinger-Fock construction and the coherent state formulation, (the relevant Perelomov coherent states were already given in \cite{Freidel:2010tt}), are left as open problems as the 
evident generalisation of the problem to other Grassmannians. 

\begin{center}
\Large{\bf Acknowledgements}
\end{center}

We would like to thank D.~O'Connor, B.~P.~Dolan and C.~S\"amann for helpful discussions at DIAS. S.~M. was supported by the Belgian Federal Office for Scientific, Technical and Cultural 
Affairs through the Interuniversity Attraction Pole P6/11. I.~H. thanks A. Wipf for useful discussions and the TPI-FSUJ for kind hospitality and continued support through the DFG grants 
Gi 328/3-2 and SFB-TR18.

\renewcommand{\theequation}{A.\arabic{equation}}
\setcounter{equation}{0}

\begin{appendix}

\section{Remainder of the Laplacian} \label{SectP}

In this appendix we prove the equation (\ref{finallaplacian}) used in the calculation of the Laplacian's spectrum. We start by giving some useful formulae used throughout the 
calculation:

\begin{eqnarray}
\Ad_{\mu\nu}\hat{L}_{\al}{}^\mu |_{\F_L}&=&\Ad_{\al\nu}\N |_{\F_L}~, \quad \label{contract1} \\
 A^{\mu\nu}\hat{L}_{\mu}{}^\al |_{\F_L}&=&(\N+n-1)A^{\al\nu}|_{\F_L}~, \label{contract2}
\end{eqnarray}
\eq
\!\!\!\!\!\! \Ad_{\al \bt}\Ad_{\mu \nu} + \Ad_{\al \mu}\Ad_{\nu \bt} + \Ad_{\al \nu} \Ad_{\bt \mu}=0 ~.
\qe
The last (Pl\"ucker) equation follows from $ \epsilon_{i[j}\epsilon_{kl]} \ad{\al}{i} \ad{\bt}{j} \ad{\mu}{k} \ad{\nu}{l} = 0$.

We would like to calculate the eigenvalues of the least trivial part of the Laplacian (\ref{laplacian}). To this end we will first consider polarization tensors with $l+p=L$

\begin{align} 
\bD_{\avec_l,\bvec_l;\gvec_p}^{\mvec_l,\nvec_l;\evec_p}=\frac{1}{\sqrt{Q_n(l,p,L)}} 
\calP_{(\avec_l,\bvec_l;\gvec_p)(\avec_l',\bvec_l';\gvec_p')}^{(\mvec_l,\nvec_l;\evec_p)(\mvec_l',\nvec_l';\evec_p')}\ket{\mvec_l'\evec'_p  ,\nvec'_l\dvec_p }
\bra{\avec'_l\gvec'_p ,\bvec'_l\dvec_p}\label{appendixD}
\end{align}
and the case $p=1$. To simplify matters, let us take an element of the matrix algebra, $\bm{\Phi}$, built from such a polarization tensor

\eq 
\bm{\Phi}=f^{\avec_L,\,\bvec_{L-1}}_{\mvec_L,\,\nvec_{L-1}} \ket{\avec_L,\,\bvec_{L-1}\g}\bra{\mvec_L,\,\nvec_{L-1}\g} ~,
\qe
where $f^{\avec_L,\,\bvec_{L-1}}_{\mvec_L,\,\nvec_{L-1}}$ vanishes under any contraction of a lower and an upper index. We are interested in the action of the non-trivial part of the 
Laplacian on $\bm{\Phi}$ and its generalisation to single contractions on $k$ composite oscillators. To this end it is convienient to write $\bm{\Phi}$ in the form

\eq 
\bm{\Phi}=:\Ad_{\mu\g}\Phi^\mu_\nu A^{\nu\g}~, 
\qe
where $\Phi^{\al_L}_{\mu_L}=\frac{1}{L(L+1)}f^{\avec_L,\,\bvec_{L-1}}_{\mvec_L,\,\nvec_{L-1}} \ket{\avec_{L-1},\,\bvec_{L-1}}\bra{\mvec_{L-1},\,\nvec_{L-1}}$~.
We then have
\begin{align}
\Lh_\al{}^\bt \bm{\Phi}\Lh_\bt{}^\al&=\Lh_\al{}^\bt\Ad_{\mu\g}\Phi^\mu_\nu A^{\nu\g}\Lh_\bt{}^\al\\
&=[\Lh_\al{}^\bt,\,\Ad_{\mu\g}]\Phi^\mu_\nu A^{\nu\g}\Lh_\bt{}^\al+\Ad_{\mu\g}\Lh_\al{}^\bt\Phi^\mu_\nu [A^{\nu\g},\,\Lh_\bt{}^\al]\\
&=\Ad_{\al\g}\Phi^\mu_\nu A^{\nu\g}\Lh_\mu{}^\al +\Ad_{\mu\al}\Phi^\mu_\nu A^{\nu\g}\Lh_\g{}^\al+\Ad_{\mu\g}\Lh_\al{}^\g\Phi^\mu_\nu A^{\nu\al} +
\Ad_{\mu\g}\Lh_\al{}^\nu\Phi^\mu_\nu A^{\al\g}
\end{align}
Observing that $\Lh_\al{}^\nu \Phi^\mu_\nu=\Phi^\mu{}_\nu\Lh_\mu{}^\al=\Lh_{\al}{}^\bt \Phi^\mu_\nu \Lh_\bt{}^\s=0$ and using the relations (\ref{contract1}) and 
(\ref{contract2}), we find
\begin{align}
\Lh_\al{}^\bt \bm{\Phi}\Lh_\bt{}^\al&=(\N + \N^\ttR+n-4)\bm{\Phi}~.
\end{align}
For generality, we have left the left and right acting number operators unevalulated so that the result is valid for non-square matrices. For the case at hand, we have 
$\N=\N^\ttR=L=l+p$.

Generalizing to $p$ contractions is not difficult. We just continue in an iterative fashion, remembering that the $p$ indices $\gvec$ (and $\tvec$) in (\ref{appendixD}) are 
symmetrized. We find
\eq \hat{L}_{\al}{}^\bt\bD_{\avec_l,\bvec_l;\gvec_p}^{\mvec_l,\nvec_l;\evec_p}\hat{L}_{\bt}{}^\al=p(\N +\N^\ttR+n-p-3)
\bD_{\avec_l,\bvec_l;\gvec_p}^{\mvec_l,\nvec_l;\evec_p}~.\qe

\renewcommand{\theequation}{B.\arabic{equation}}
\setcounter{equation}{0}

\setcounter{section}{2}

\subsection{Factor $Q_n(l,p,l+p)$} \label{SectQ}

We begin by writing the highest weight vector in the representation $(p,l,0,\cdots,0,l,p)$, it is given by the polarization tensor corresponding to the Young diagram

{\Large$$
\mbox{ \raisebox{0.34cm}{$\overbrace{\overline{\young(\n\Cdots\n)}}^p$}  } \hspace{-1.98mm}
\mbox{\raisebox{0.34cm}{$\overbrace{\overline{\young(\n\Cdots\n)}}^l $}} \hspace{-2.03cm} 
\mbox{\raisebox{-.34cm}{$\young(\nmo\Cdots) \hspace{-.1mm} \young(\nmo)$}}  \hspace{-1.1mm} 
\overbrace{\young(\one\Cdots\one,\two\Cdots\two)}^l \hspace{-1.1mm}
\mbox{\raisebox{0.34cm}{$\overbrace{\young(\one\Cdots\one)}^p$} }
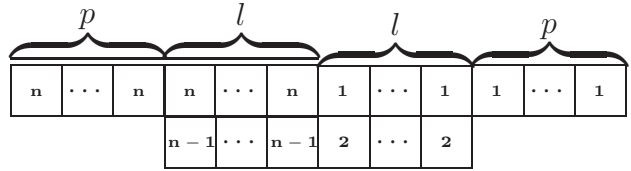
$$}

and explicitly

\eq 
\bD_0 = \frac{1}{\sqrt{Q_n (l,p,l+p)}} | \underline{\bf 1} \overline{ \bf 1}, \underline{ \bf 2} \overline{\dvec} \rangle \langle \underline{\bf n} \overline{\bf n}, 
\underline{\bf n-1} \overline{\dvec}|
\qe
where the notation $\underline{\bf x} = x_1, x_2, \cdots, x_l$ and $\overline{\bf y} = y_1, y_2, \cdots , y_p$ was  introduced for convenience. We then demand that the norm of the highest 
weight vector be unity

\eq
1=Tr(\bD_0^{\dagger}\bD_0)= \frac{1}{Q_n(l,p,l+p)}  \langle \underline{\bf n} \overline{\bf n}, \underline{\bf n-1} \overline{\dvec} | \underline{\bf n} \overline{\bf n}, 
\underline{\bf n-1} \overline{\ovec} \rangle \langle \underline{\bf 1} \overline{\bf 1}, \underline{\bf 2} \overline{\ovec} |\underline{\bf 1} \overline{\bf 1}, \underline{\bf 2} 
\overline{\dvec}  \rangle
\qe
hence,

\eq \label{D1}
Q_n (l,p,l+p) = \langle \underline{\bf n} \overline{\bf n}, \underline{\bf n-1} \overline{\dvec} | \underline{\bf n} \overline{\bf n}, \underline{\bf n-1} \overline{\ovec} \rangle 
\langle \underline{\bf 1} \overline{\bf 1}, \underline{\bf 2} \overline{\ovec} |\underline{\bf 1} \overline{\bf 1}, \underline{\bf 2} \overline{\dvec}  \rangle
\qe
The following notation is also useful:

\begin{eqnarray}\non
\theta({\bf \underline{x}}) & = & \prod_{k=1}^l \theta (x_k),\quad \quad \theta(x) = 1-\delta_{x}^{1} \\ \non
\lambda({\bf \underline{x}}) & = & \prod_{k=1}^l \lambda (x_k), \quad \quad \lambda(x) = 1-\delta_{x}^{n}
\end{eqnarray}
We observe the identity

\eq \non
\langle  \underline{\bf 1} , \underline{\mvec} | \underline{\bf 1}, \underline{\nvec} \rangle = 
\theta(\underline{\mvec})\theta(\underline{\nvec}) \langle  \underline{\bf 1} , \underline{\mvec} | \underline{\bf 1}, \underline{\nvec} \rangle.
\qe
Here it is understood that no summation is carried out when the repeated indices appear inside $\theta$ or $\lambda$. Using this identity we may compute easily

\eq \label{D2}
\langle  \underline{\bf 1} , \underline{\mvec} | \underline{\bf 1}, \underline{\nvec} \rangle = \frac{1}{2^l} \mathcal{S}^{\underline{\mvec}}_{\underline{\nvec}} ~
\theta(\underline{\mvec})\theta(\underline{\nvec})
\qe
Where $\mathcal{S}$ denotes the projector onto totally symmetrized tensors. Using (\ref{D2}) the equation (\ref{D1}) can now be written as

\begin{eqnarray} \label{D2.5} \non
Q_n(l,p,l+p) &=& \frac{1}{4^{l+p}} \mathcal{S}^{\underline{\bf 2} \overline{\ovec}}_{\underline{\bf 2} \overline{\dvec}} 
\mathcal{S}^{\underline{\bf n-1} \overline{\dvec}}_{ \underline{\bf n-1} \overline{\ovec}}~ \theta(\overline{\dvec})\theta(\overline{\ovec}) 
\lambda(\overline{\dvec})\lambda(\overline{\ovec}) \\ 
             &=& \frac{1}{4^{l+p}} \mathcal{S}^{\underline{\bf 2} \overline{\Avec}}_{\underline{\bf 2} \overline{\Bvec}}  
\mathcal{S}^{\underline{\bf n-1} \overline{\Bvec}}_{ \underline{\bf n-1} \overline{\Avec}}
\end{eqnarray}
where the indices $A_i$ or $B_j$ run only in the interval $2,\cdots, n-1$ and, in this section only, $i,j=1, \cdots, p$. We evaluate this expression by cases: The case $n=3$ is 
trivial since all indices $A_i,B_j$ can only take the value $2$. In this case one has $Q_{3}(l,p,l+p) = (1/4)^{l+p}$. When $n \ge 4$ to evaluate this factor we expand the first 
projector in the following manner:

\eq \label{D3}
\mathcal{S}^{\underline{\bf 2} \overline{\Avec}}_{\underline{\bf 2} \overline{\Bvec}}  = \frac{l!p!}{(l+p)!}\sum_{k=0}^{min\{l,p\}}\frac{1}{k!}\binom{l}{k} 
\sum_{\begin{subarray}{1} i_1, \cdots, i_k =1 \\ ~~~\mbox{\scriptsize a.d.} \end{subarray}}^p \!\!\!   
\delta_2^{A_{i_1}}\cdots \delta_2^{A_{i_k}} \mathcal{S}_{\overline{\Bvec}}^{\overline{\Avec}_{i_1 \cdots i_k}\overbrace{\mbox{\scriptsize{$2 \cdots 2$}}}^{k}}
\qe
Where we introduced the notation $\overline{\Avec}_{i_1 \cdots i_k}$ to denote the string of indices that results by removing $A_{i_1}, A_{i_2}, \cdots, A_{i_k}$ from 
$\overline{\Avec}$. The notation ``a.d.'' in the sum means that the sum is to be taken over ``all distinct'' values of their indices. This formula is proven in the next subsection.

Substituting (\ref{D3}) into (\ref{D2.5}) gives after reordering terms:

\begin{eqnarray} \non
Q_{n \ge 4}(l,p,l+p) &=& \frac{l! p!}{4^{l+p} (l+p)!} \sum_{k=0}^{ min\{l, p\} } \sum_{\begin{subarray}{1} i_1, \cdots, i_k =1 \\~~~ \mbox{\scriptsize a.d} \end{subarray}}^p 
\frac{1}{k!} \binom{l}{k} 
\mathcal{S}^{\underline{\bf n-1} \overbrace{\mbox{\scriptsize{$2\cdots 2$} }  }^k \overline{\Avec}_{1,\cdots,k} }_{\underline{\bf n-1} 
\underbrace{\mbox{\scriptsize{$2\cdots 2$}} }_k \overline{\Avec}_{1,\cdots,k}} \\ \non
&=& \left( \frac{l!}{2^{l+p} (l+p)!} \right)^2 p! (l+n+p-3)! \sum_{k=0}^{min \{l,p\}} \binom{l}{k} \binom{p}{k} \frac{k!}{(l+n+k-3)!} \\ \label{Qminimal}
&=&\left( \frac{l!p!}{2^{l+p} (l+p)!} \right)^2 ~~\sum_{k=0}^{min \{l,p\}} \binom{l}{k} \binom{l+p+n-3}{p-k} \\ \non
&=&\left( \frac{l!p!}{2^{l+p} (l+p)!} \right)^2  \binom{2l+p+n-3}{p}
\end{eqnarray}

\subsection{Proof of an identity}

Here we present the proof by induction of the identity (\ref{D3}) considering separately the cases $p \geq l$ and $p \leq l$. First of all, the case $l=p=0$ is trivial.
It is very easy to check that (\ref{D3}) is valid for $p \geq l=0$, as it is to see it remains valid when $p \geq l = 1$.  Let us now suppose first that $p > l \geq 1$ 
and assume that (\ref{D3}) is true for such values $p$ and $l$, assume also that $p \geq l+1$, we shall prove that the formula then holds for the values $p,l+1$. 

Consider then the projector with $p,l+1$ and expand it as

\eq \label{D4.5}
\mathcal{S}^{\underline{\bf 2}2 \overline{\Avec}}_{\underline{\bf 2}2 \overline{\Bvec}} = \frac{1}{l+p+1} \Big( (l+1) 
\mathcal{S}^{\underline{\bf 2} \overline{\Avec}}_{\underline{\bf 2} \overline{\Bvec}} + \sum_{q=1}^p \delta^{A_q}_2 
\mathcal{S}^{\underline{\bf 2} 2 \overline{\Avec}_q }_{\underline{\bf 2} \overline{\Bvec}} \Big)
\qe
our attention is drawn to the second term in the r.h.s., we define the indices $A'_k = A_k$ for $k \neq q$ and $A'_q = 2$; using the symmetry of $\mathcal{S}$ we rewrite this term as 

\eq \non 
\frac{1}{l+p+1}\sum_{q=1}^p \delta^{A_q}_2 \mathcal{S}^{\underline{\bf 2} \overline{\Avec'}}_{\underline{\bf 2} \overline{\Bvec}}
\qe
Next we use (\ref{D3}), which is assumed to be valid for $p,l$, to expand this term, notice that $l=min\{l,p\}$. We make use of the following partition of the sum involved:

\eq \label{D5}
\sum_{\begin{subarray}{1} i_1, \cdots, i_k =1 \\ ~~~\mbox{\scriptsize a.d.} \end{subarray}}^p f^{i_1 \cdots i_k} =
\sum_{\begin{subarray}{1} i_1, \cdots, i_k =1 \\ ~\mbox{\scriptsize a.d., }\neq q \end{subarray}}^p f^{i_1 \cdots i_k} + 
\sum_{r=1}^k \sum_{\begin{subarray}{1} ~~i_1, \cdots \hat{i}_r \cdots, i_k =1 \\ ~~~\mbox{\scriptsize a.d., }\neq q \end{subarray}}^p 
f^{i_1 \cdots \mbox{\scriptsize{ $\overbrace{q}^r$}}   \cdots i_k}
\qe
valid for an arbitrary indexed object $f$. After this decomposition (\ref{D3}) is the sum of two terms, namely

\begin{eqnarray} \non
 \frac{l! p!}{(l+p+1)!}\sum_{k=0}^l \frac{1}{k!} \binom{l}{k} \sum_{\begin{subarray}{1} i_1, \cdots, i_{k+1} =1 \\ ~~~\mbox{\scriptsize a.d.} \end{subarray}}^p    
\delta^{A_{i_1}}_{2} \cdots \delta^{A_{i_{k+1}}}_{2} \mathcal{S}_{\overline{\Bvec}}^{\overline{\Avec}_{i_1 \cdots i_{k+1}}\overbrace{\mbox{\scriptsize{$2 \cdots 2$}}}^{k+1}}
\\ \! \! \! \! \! \!\! \! \! \! \! \non +\quad  \frac{l! p!}{(l+p+1)!} \sum_{k=1}^{l} \frac{1}{(k-1)!} \binom{l}{k}
\sum_{\begin{subarray}{1} i_1, \cdots, i_{k} =1 \\ ~~~\mbox{\scriptsize a.d.} \end{subarray}}^p \delta^{A_{i_1}}_{2} \cdots \delta^{A_{i_{k}}}_{2} 
\mathcal{S}_{\overline{\Bvec}}^{\overline{\Avec}_{i_1 \cdots i_{k}}\overbrace{\mbox{\scriptsize{$2 \cdots 2$}}}^{k}}
\end{eqnarray}
in the first term above the renaming $q \to i_{k+1}$ was made and the condition $p\geq l+1$ tacitly used, in the second term care must be taken with the term $k=0$ which vanishes 
already in (\ref{D5}). The renaming $i_1 \cdots \hat{i}_r \cdots i_k q \to i_1 \cdots i_k$ was carried out in the second term above. 
By use of elementary identities we may combine both terms above into 

\eq \label{D6}
\frac{l! p!}{(l+p+1)!} \sum_{k=1}^{l+1} \frac{1}{(k-1)!} \binom{l+1}{k}\sum_{\begin{subarray}{1} i_1, \cdots, i_{k} =1 \\ ~~~\mbox{\scriptsize a.d.} \end{subarray}}^p
\delta^{A_{i_1}}_{2} \cdots \delta^{A_{i_{k}}}_{2} \mathcal{S}_{\overline{\Bvec}}^{\overline{\Avec}_{i_1 \cdots i_{k}}\overbrace{\mbox{\scriptsize{$2 \cdots 2$}}}^{k}}
\qe
As a final step we substitute (\ref{D3}) into the first term of the r.h.s. in (\ref{D4.5}), combining the resulting term with (\ref{D6}) gives, after a little manipulation

\eq \non
\mathcal{S}^{\underline{\bf 2}2 \overline{\Avec}}_{\underline{\bf 2}2 \overline{\Bvec}} = \frac{(l+1)!p!}{(l+p+1)!}\sum_{k=0}^{l+1}\frac{1}{k!}\binom{l+1}{k} 
\sum_{\begin{subarray}{1} i_1, \cdots, i_k =1 \\ ~~~\mbox{\scriptsize a.d.} \end{subarray}}^p \!\!\!   
\delta_2^{A_{i_1}}\cdots \delta_2^{A_{i_k}} \mathcal{S}_{\overline{\Bvec}}^{\overline{\Avec}_{i_1 \cdots i_k}\overbrace{\mbox{\scriptsize{$2 \cdots 2$}}}^{k}}
\qe 
this ends the proof by induction over $l$ for all the cases considered above. In what follows we prove the cases $p \leq l$ by induction on $p$. To this end assume that (\ref{D3}) is 
valid for the values $l,p$ such that $p + 1 \leq l$ and $1 \leq p < l$ hold and prove then its validity for $l,p+1$. The identity (\ref{D3}) is trivially true for 
$l \geq p=0$, it can also be readily verified that it is also satisfied for $l \geq p =1$. Notice that now $p=min\{l,p \}$ is to be used. As a first step we expand the projector

\eq \label{D7}
\mathcal{S}^{\underline{\bf 2} \overline{\Avec} A_{p+1}}_{\underline{\bf 2} \overline{\Bvec}B_{p+1}} = \frac{1}{(l+p+1)}\Big( l \delta_{2}^{A_{p+1}}
\mathcal{S}^{\underline{\bf 2} \overline{\Avec}}_{\underline{\bf 2}_1 \overline{\Bvec}B_{p+1}}  + \sum_{h=1}^{p+1}\delta^{A_{p+1}}_{B_h} 
\mathcal{S}^{\underline{\bf 2} \overline{\Avec}}_{\underline{\bf 2}(\overline{\Bvec}B_{p+1})_h} \Big) 
\qe  
Where the notation $\underline{\bf 2}_1 = \overbrace{2\cdots 2}^{l-1}$ and $(\overline{\Bvec}B_{p+1})_h = B_1 \cdots\hat{B}_h \cdots B_{p+1}$ has been introduced following previous 
conventions. Using our decomposition (\ref{D3}), valid for $l,p$, in the second term of the r.h.s in (\ref{D7}) we obtain 

\eq \label{D7.5}
\frac{l!p!}{(l+p+1)!} \sum_{h=1}^{p+1} \delta^{A_{p+1}}_{B_h} \sum_{k=0}^p \frac{1}{k!} \binom{l}{k} \sum_{\begin{subarray}{1} i_1, \cdots, i_k =1 \\ ~~~\mbox{\scriptsize a.d.} 
\end{subarray}}^p \!\!\!   \delta_2^{A_{i_1}}\cdots \delta_2^{A_{i_k}} 
\mathcal{S}_{(\overline{\Bvec}B_{p+1})_h}^{\overline{\Avec}_{i_1 \cdots i_k}\overbrace{\mbox{\scriptsize{$2 \cdots 2$}}}^{k}}
\qe

Now we decompose the first term in the r.h.s. of (\ref{D7}) as 

\eq \label{D8}
\frac{l}{(l+p+1)(l+p)} \delta^{A_{p+1}}_2 \Big( l \delta^{2}_{B_{p+1}} \mathcal{S}^{\underline{\bf 2}_1 \overline{\Avec}}_{\underline{\bf 2}_1 \overline{\Bvec}} 
+ \sum_{q=1}^p \delta^{A_q}_{B_{p+1}}\mathcal{S}^{\underline{\bf 2}_1 \overline{\Avec}'}_{\underline{\bf 2}_1 \overline{\Bvec}}   \Big)
\qe
once again use (\ref{D3}) to decompose the first term in (\ref{D8}), we find using also $p+1 \leq l$

\eq \label{D8.5}
\frac{l!p! l}{(l+p+1)!} \delta^{A_{p+1}}_{2} \delta^{2}_{B_{p+1}} \sum_{k=0}^{p} \frac{1}{k!}\binom{l-1}{k}\sum_{\begin{subarray}{1} i_1, \cdots, i_k =1 \\ ~~~\mbox{\scriptsize a.d.} 
\end{subarray}}^p \!\!\!   \delta_2^{A_{i_1}}\cdots \delta_2^{A_{i_k}} 
\mathcal{S}_{\overline{\Bvec}}^{\overline{\Avec}_{i_1 \cdots i_k}\overbrace{\mbox{\scriptsize{$2 \cdots 2$}}}^{k}}
\qe
For the second term in (\ref{D8}) we use (\ref{D3}) and (\ref{D5}), after simplifying the two terms obtained we get

\begin{eqnarray} \non
\frac{l! p!}{(l+p+1)!} \delta^{A_{p+1}}_2 \sum_{k=1}^p \frac{1}{(k-1)!}\binom{l-1}{k-1}\sum_{\begin{subarray}{1} i_1, \cdots, i_k =1 \\ ~~~\mbox{\scriptsize a.d.} 
\end{subarray}}^p \!\!\!   \delta_2^{A_{i_1}}\cdots \delta_2^{A_{i_{k-1}}}\delta^{A_{i_k}}_{B_{p+1}} 
\mathcal{S}_{\overline{\Bvec}}^{\overline{\Avec}_{i_1 \cdots i_k}\overbrace{\mbox{\scriptsize{$2 \cdots 2$}}}^{k}} \\ \non
+ ~\frac{l! p!}{(l+p+1)!} \delta^{A_{p+1}}_2 \sum_{k=1}^p \frac{1}{(k-1)!}\binom{l-1}{k}\sum_{\begin{subarray}{1} i_1, \cdots, i_k =1 \\ ~~~\mbox{\scriptsize a.d.} 
\end{subarray}}^p \!\!\!   \delta_2^{A_{i_1}}\cdots \delta_2^{A_{i_{k-1}}}\delta^{A_{i_k}}_{B_{p+1}} 
\mathcal{S}_{\overline{\Bvec}}^{\overline{\Avec}_{i_1 \cdots i_k}\overbrace{\mbox{\scriptsize{$2 \cdots 2$}}}^{k}}
\end{eqnarray}
in the first term care has to be taken as the contribution with $k=p$ does not appear in the sum, we renamed $q \to i_{k+1}$ and later made $k \to k-1$ within this sum recovering thus 
the upper limit of the sum, $p$. In the second term the contribution $k=0$ is not present and the relabeling $i_1 \cdots \hat{i}_r \cdots i_k q \to i_1 \cdots i_k$ was made. It is 
easy to recombine these two contributions, so the second term in (\ref{D8}) simplifies down to

\eq \label{D9}
\frac{l! p!}{(l+p+1)!} \delta^{A_{p+1}}_2 \sum_{k=1}^p \frac{1}{(k-1)!}\binom{l}{k}\sum_{\begin{subarray}{1} i_1, \cdots, i_k =1 \\ ~~~\mbox{\scriptsize a.d.} 
\end{subarray}}^p \!\!\!   \delta_2^{A_{i_1}}\cdots \delta_2^{A_{i_{k-1}}}\delta^{A_{i_k}}_{B_{p+1}} 
\mathcal{S}_{\overline{\Bvec}}^{\overline{\Avec}_{i_1 \cdots i_k}\overbrace{\mbox{\scriptsize{$2 \cdots 2$}}}^{k}}
\qe
We have thus shown that (\ref{D7}) = (\ref{D7.5}) + (\ref{D8.5}) + (\ref{D9}). To proceed further we will compare (\ref{D7}) with the expansion

\eq \label{D10}
\mathcal{S}^{\underline{\bf 2} \overline{\Avec}A_{p+1}}_{\underline{\bf 2} \overline{\Bvec}B_{p+1}}  = \frac{l!(p+1)!}{(l+p+1)!}\sum_{k=0}^{p+1}\frac{1}{k!}\binom{l}{k} 
\sum_{\begin{subarray}{1} i_1, \cdots, i_k =1 \\ ~~~\mbox{\scriptsize a.d.} \end{subarray}}^{p+1} \!\!\!   
\delta_2^{A_{i_1}}\cdots \delta_2^{A_{i_k}} \mathcal{S}_{\overline{\Bvec}B_{p+1}}^{(\overline{\Avec}A_{p+1})_{i_1 \cdots i_k}\overbrace{\mbox{\scriptsize{$2 \cdots 2$}}}^{k}}
\qe
notice that $p+1= min\{l,p+1\}$ under our assumption. Expanding the sum in the r.h.s of (\ref{D10}) using (\ref{D5}) with $q=p+1$, and after the relabeling 
$i_1 \cdots \hat{i}_r \cdots i_k \to i_1 \cdots i_{k-1} $ in the second term thus obtained, yields

\begin{eqnarray} \label{D11}
&&\frac{l! (p+1)!}{(l+p+1)!}\sum_{k=0}^{p} \frac{1}{k!} \binom{l}{k} \sum_{\begin{subarray}{1} i_1, \cdots, i_k =1 \\ ~~~\mbox{\scriptsize a.d.} 
\end{subarray}}^p \!\!\!   \delta_2^{A_{i_1}}\cdots \delta_2^{A_{i_k}} 
\mathcal{S}_{\overline{\Bvec}B_{p+1}}^{\overline{\Avec}_{i_1 \cdots i_k}A_{p+1}\overbrace{\mbox{\scriptsize{$2 \cdots 2$}}}^{k}} \\ \label{D12}
&+&\frac{l! (p+1)!}{(l+p+1)!} \delta^{A_{p+1}}_2 \sum_{k=1}^{p+1} \frac{1}{(k-1)!} \binom{l}{k} \sum_{\begin{subarray}{1} i_1, \cdots, i_{k-1} =1 \\ ~~~\mbox{\scriptsize a.d.} 
\end{subarray}}^p \!\!\!   \delta_2^{A_{i_1}}\cdots \delta_2^{A_{i_{k-1}}} 
\mathcal{S}_{\overline{\Bvec}B_{p+1}}^{\overline{\Avec}_{i_1 \cdots i_{k-1}}\overbrace{\mbox{\scriptsize{$2 \cdots 2$}}}^{k}}
\end{eqnarray}
we further noticed that the term $k=p+1$ does not contribute for (\ref{D11}) (due to the restrictions of the sum over $i$'s) and neither does the term $k=0$ in (\ref{D12}). It is not 
difficult to realise that (\ref{D11}) is identical to (\ref{D7.5}). Next make $k \to k+1$ in (\ref{D12}) and expand to cast it into the form

\begin{eqnarray} \non
\frac{l!p!}{(l+p+1)!} \delta^{A_{p+1}}_2 \sum_{k=0}^p \frac{1}{k!} \binom{l}{k+1} \sum_{\begin{subarray}{1} i_1, \cdots, i_{k} =1 \\ ~~~\mbox{\scriptsize a.d.} 
\end{subarray}}^p \!\!\!   \delta_2^{A_{i_1}}\cdots \delta_2^{A_{i_{k}}} \times \\ \label{D13}
 \Big( \sum_{\begin{subarray}{1} ~~~r=1 \\ r \neq i_1, \cdots, i_k \end{subarray} }^p \delta^{A_r}_{B_{p+1}} 
\mathcal{S}_{\overline{\Bvec}}^{\overline{\Avec}_{i_1 \cdots i_{k}r}\overbrace{\mbox{\scriptsize{$2 \cdots 2$}}}^{k+1}} + 
(k+1)\delta^2_{B_{p+1}}\mathcal{S}_{\overline{\Bvec}}^{\overline{\Avec}_{i_1 \cdots i_{k}}\overbrace{\mbox{\scriptsize{$2 \cdots 2$}}}^{k}} \Big)
\end{eqnarray}
A direct computation shows that the second term in (\ref{D13}) is in fact equal to (\ref{D8.5}). For the first term above we rename $r \to i_{k+1}$, observe that once again the term 
with $k=p$ is not present and then make $k\to k-1$ to transform it directly into (\ref{D9}). Therefore we have shown that (\ref{D7}) = (\ref{D10}), this concludes the proof by 
induction. The formula (\ref{D3}) has been established for any integers $l,p \geq 0$.

\end{appendix}

\bibliographystyle{JHEP}

\bibliographystyle{abbrv}

\end{document}